\documentclass[reqno,a4paper,11pt]{article}
\pdfoutput=1
\usepackage{xcolor}

\usepackage{graphicx}
\usepackage[textwidth = 430 pt, textheight = 630 pt]{geometry}

\definecolor{MyDarkBlue}{rgb}{0.15,0.25,0.45}
\usepackage{epsfig,rotating}
\usepackage{amsmath,amssymb}
\usepackage{amsfonts}
\usepackage{mathrsfs}
\usepackage{bbm}
\usepackage[normalem]{ulem}

\usepackage{latexsym}
\usepackage{amsthm}
\usepackage[all,knot]{xy}
\xyoption{arc}

\usepackage[utf8x]{inputenc}

\usepackage{hyperref}
\hypersetup{
hypertexnames=false,
colorlinks=true,
citecolor=MyDarkBlue,
linkcolor=MyDarkBlue,
urlcolor=MyDarkBlue,
pdfauthor={Patricia Ritter, Christian S\"amann and Lennart Schmidt},
pdftitle={Generalized Higher Gauge Theory},
pdfsubject={hep-th math-ph}
breaklinks=true
}

\usepackage{tikz}
\usepackage{mathtools}

\let\fn\footnote
\renewcommand{\footnote}[1]{\linespread{1.1}\fn{#1}\linespread{1.29}}

\makeatletter\renewcommand{\section}{\@startsection
{section}{1}{\z@}{-3.5ex plus -1ex minus
    -.2ex}{2.3ex plus .2ex}{\bf }}
\makeatletter\renewcommand{\subsection}{\@startsection{subsection}{2}{\z@}{-3.25ex
plus -1ex minus
   -.2ex}{1.5ex plus .2ex}{\bf }}
\makeatletter\renewcommand{\subsubsection}{\@startsection{subsubsection}{3}{-2.45ex}{-3.25ex
plus -1ex minus -.2ex}{1.5ex plus .2ex}{\it }}
\renewcommand{\thesection}{\arabic{section}}
\renewcommand{\thesubsection}{\arabic{section}.\arabic{subsection}}
\renewcommand{\@seccntformat}[1]{\@nameuse{the#1}.~~}

\renewcommand{\theequation}{\thesection.\arabic{equation}}
\makeatletter \@addtoreset{equation}{section}

\usepackage[toc,page]{appendix}

\newtheorem{thm}{Theorem}[section]
\renewcommand{\thethm}{\thesection.\arabic{thm}}

\newtheorem{remark}[thm]{Remark}

\renewcommand{\appendices}{
\section*{Appendix}\label{appendices}\setcounter{subsection}{0}
\addcontentsline{toc}{section}{Appendix}
\setcounter{equation}{0}
\makeatletter
\renewcommand{\theequation}{\Alph{subsection}.\arabic{equation}}
\renewcommand{\thesubsection}{\Alph{subsection}}
\renewcommand{\thethm}{\Alph{subsection}.\arabic{thm}}
\@addtoreset{equation}{subsection}
\@addtoreset{thm}{subsection}
\makeatother
}

\def\slasha#1{\setbox0=\hbox{$#1$}#1\hskip-\wd0\hbox to\wd0{\hss\sl/\/\hss}}

\def\periodb#1{\setbox0=\hbox{$#1$}#1\hskip-\wd0\hbox to\wd0{-}}

\newcommand{\lbr}{(\hspace{-0.1cm}(}
\newcommand{\rbr}{)\hspace{-0.1cm})}

\newcommand{\unit}{\mathbbm{1}}   			
\newcommand{\id}{\mathrm{id}}   			

\newcommand{\CA}{\mathcal{A}}    			

\newcommand{\CB}{\mathcal{B}}

\newcommand{\CC}{\mathcal{C}}

\newcommand{\CF}{\mathcal{F}}

\newcommand{\CG}{\mathcal{G}}

\newcommand{\CL}{\mathcal{L}}
\newcommand{\CM}{\mathcal{M}}

\newcommand{\CN}{\mathcal{N}}
\newcommand{\CO}{\mathcal{O}}

\newcommand{\CS}{\mathcal{S}}

\newcommand{\CV}{\mathcal{V}}

\newcommand{\CX}{\mathcal{X}}

\newcommand{\fra}{\mathfrak{a}}				
\newcommand{\frg}{\mathfrak{g}}				

\newcommand{\Walg}{\mathrm{W}}	
\newcommand{\CEalg}{\mathrm{CE}}

\newcommand{\FR}{\mathbbm{R}}     			
\newcommand{\NN}{\mathbbm{N}}     			

\newcommand{\dd}{\mathrm{d}}     			
\newcommand{\dpar}{\partial}     			
\newcommand{\embd}{{\hookrightarrow}}     		

\newcommand{\di}{\mathrm{i}}     			
\newcommand{\eps}{{\varepsilon}}			

\newcommand{\eand}{{\qquad\mbox{and}\qquad}}     		
\newcommand{\ewith}{{\qquad\mbox{with}\qquad}}
\newcommand{\efor}{{\qquad\mbox{for}\qquad}}

\newcommand{\der}[1]{\frac{\dpar}{\dpar #1}}   		

\newcommand{\aso}{\mathfrak{so}}

\newcommand{\sG}{\mathsf{G}}
\newcommand{\sL}{\mathsf{L}}

\newcommand{\sLie}{\mathsf{Lie}}

\newcommand{\sH}{\mathsf{H}}

\newcommand{\CatMan}{\mathsf{Man}^\infty}

\newcommand{\acton}{\vartriangleright}     			
\renewcommand{\remark}[1]{}     				
\def\tyng(#1){\hbox{\tiny$\yng(#1)$}}			
\def\tyoung(#1){\hbox{\tiny$\young(#1)$}}			

\newcommand{\beq}{\begin{eqnarray}}
\newcommand{\eeq}{\end{eqnarray}}

\newcommand{\sSym}{{\sf Sym}}

\newcommand{\sfs}{{\sf s}}
\newcommand{\sft}{{\sf t}}

\begin{document}
\begin{titlepage}
\begin{flushright}
 DIFA 2015\\
 EMPG--15--26
\end{flushright}
\vskip 2.0cm
\begin{center}
{\LARGE \bf Generalized Higher Gauge Theory}
\vskip 1.5cm
{\Large Patricia Ritter$^a$, Christian S\"amann$^b$ and Lennart Schmidt$^b$}
\setcounter{footnote}{0}
\renewcommand{\thefootnote}{\arabic{thefootnote}}
\vskip 1cm
{\em${}^a$ Dipartimento di Fisica ed Astronomia \\
Universit\`a di Bologna and INFN, Sezione di Bologna\\
Via Irnerio 46, I-40126 Bologna, Italy
}\\[0.5cm]
{\em ${}^b$ Maxwell Institute for Mathematical Sciences\\
Department of Mathematics, Heriot-Watt University\\
Colin Maclaurin Building, Riccarton, Edinburgh EH14 4AS,
U.K.}\\[0.5cm]
{Email: {\ttfamily patricia.ritter@bo.infn.it , C.Saemann@hw.ac.uk , ls27@hw.ac.uk}}
\end{center}
\vskip 1.0cm
\begin{center}
{\bf Abstract}
\end{center}
\begin{quote}
We study a generalization of higher gauge theory which makes use of generalized geometry and seems to be closely related to double field theory. The local kinematical data of this theory is captured by morphisms of graded manifolds between the canonical exact Courant Lie 2-algebroid $TM\oplus T^*M$ over some manifold $M$ and a semistrict gauge Lie 2-algebra. We discuss generalized curvatures and infinitesimal gauge transformations. Finite gauge transformation as well as global kinematical data are then obtained from principal 2-bundles over 2-spaces. As dynamical principle, we consider first the canonical Chern-Simons action for such a gauge theory. We then show that a previously proposed 3-Lie algebra model for the six-dimensional (2,0) theory is very naturally interpreted as a generalized higher gauge theory.
\end{quote}
\end{titlepage}

\tableofcontents

\section{Introduction and results}

Higher gauge theory describes the parallel transport of extended objects transforming under local internal symmetries. There are well-known no-go theorems stating that in a naive setting, the internal symmetry group has to be abelian for objects with positive dimension. To avoid these theorems, one has to categorify the ingredients of usual gauge theory, see e.g.\ \cite{Baez:2010ya} for details. This leads in particular to categorified structure groups, known as $n$-groups, as well as categorified notions of principal bundles known as principal $n$-bundles. 

One severe open problem in higher gauge theory is the lack of non-trivial examples of non-abelian principal $n$-bundles with connection. For example, one would expect categorified analogues of non-abelian monopoles and instantons to exist. Although higher analogues of the twistor descriptions of monopoles and instantons have been constructed \cite{Saemann:2012uq,Saemann:2013pca,Jurco:2014mva}, the known solutions, e.g. those of \cite{Palmer:2013haa}, do not quite fit the picture.\footnote{If higher gauge theory is to describe the parallel transport of some extended objects, then a condition needs to be imposed on the curvature of the principal $n$-bundle to ensure that the parallel transport is invariant under reparameterizations. While the solutions of \cite{Palmer:2013haa} do not directly satisfy this curvature condition, one could argue that at least in the case of the self-dual strings considered in \cite{Palmer:2013haa}, the underlying parallel transport of strings is trivial and the fake curvature condition becomes physically irrelevant.}

This lack of examples presents an obstacle to both mathematical as well as physical progress in the study of higher gauge theory. It is therefore important to find generalized formulations which allow for interesting examples. In this paper, we study the case in which also the base manifold is categorified to what has been called a 2-space \cite{Baez:2004in}: a category internal to the category of smooth manifolds. 

In recent developments in string theory, there are many pointers towards the necessity of using 2-spaces instead of ordinary space-time manifolds, in particular in relation with generalized geometry and double field theory. In both contexts, it is usually the exact Courant algebroid $TM\oplus T^*M$ over some manifold $M$, which is used to give expressions a coordinate-invariant meaning, see e.g.\ \cite{Vaisman:2012ke,Hohm:2012mf,Deser:2014mxa}. It is therefore only natural to ask whether a definition of gauge theory involving this algebroid has some interesting features.

Recall that the local kinematical data of ordinary gauge theory over some manifold $M$ can be described by a morphism of graded algebras between the Chevalley-Eilenberg algebra of the gauge Lie algebra and the Weil algebra of the manifold $M$, which is the Chevalley-Eilenberg algebra of the tangent Lie algebroid $TM$. For higher gauge theory, the domain of this morphism is extended to the Chevalley-Eilenberg algebra of some $L_\infty$-algebra. In this paper, we also generalize the range of this morphism to the Chevalley-Eilenberg algebra of the Courant algebroid, cf.\ figure \ref{fig:1}. The latter should more properly be regarded as a symplectic Lie 2-algebroid, and thus we arrive at a notion of gauge theory over the 2-space $T^*M\rightrightarrows M$. 
\begin{figure}\label{fig:1}
 \begin{equation*}
 \begin{aligned}
  {\rm W}(M)={\rm CE}(TM)~~&\xleftarrow{~~~~a^*~~~~}~~{\rm CE}(\mbox{gauge Lie algebra})\\
  {\rm W}(M)={\rm CE}(TM)~~&\xleftarrow{~~~~a^*~~~~}~~{\rm CE}(\mbox{gauge $L_\infty$-algebra})\\
  {\rm W}(T^*M)={\rm CE}(T(T^*M))~~&\xleftarrow{~~~~a^*~~~~}~~{\rm CE}(\mbox{gauge $L_\infty$-algebra})\\
 \end{aligned}
 \end{equation*}
 \caption{The descriptions of local kinematical data of gauge theory, higher gauge theory and generalized higher gauge theory by morphisms $a^*$ of graded algebras. The abbreviations ${\rm CE}(X)$ and ${\rm W}(X)$ stand for the Chevalley-Eilenberg and Weil algebras of $X$, respectively.}
\end{figure}

We discuss in detail the case where the gauge $L_\infty$-algebra consists of two terms, corresponding to a semistrict Lie 2-algebra. In particular, we derive the form of the gauge potential and its curvature, which are encoded in the morphism of graded algebras and its failure to be a morphism of {\em differential} graded algebras. We also give the relevant formulas for infinitesimal gauge transformations. As we show, these results can also be obtained from the homotopy Maurer-Cartan equations of an $L_\infty$-algebra consisting of the tensor product of the gauge $L_\infty$-algebra with the Weil algebra of $T^*M$.

To glue together local kinematical data to global ones, we need a generalized principal 2-bundle structure as well as finite gauge transformations. We find both by considering principal 2-bundles over the 2-space $T^*M$.  We thus arrive at an explicit formulation of the first generalized higher Deligne cohomology class, encoding equivalence classes of these higher bundles with connection.

In a second part, we discuss two possible dynamical principles for the generalized higher connections. The first one is a Chern-Simons action, which is obtained via a straightforward generalization of the AKSZ procedure. The second one is a previously proposed set of equations for a 3-Lie algebra\footnote{not to be confused with a Lie 3-algebra}-valued $(2,0)$-tensor supermultiplet in six dimensions \cite{Lambert:2010wm}. We show that these equations find a very natural interpretation within generalized higher gauge theory. In particular, the 3-Lie algebra valued vector field featuring crucially in the equations is part of a generalized higher connection.

Among the open questions we intend to study in future work are the following. First, an additional gauge algebra-valued vector field seems to be desirable in many open questions related to the six-dimensional (2,0)-theory. It would be interesting to see if such problems can be addressed within our framework. Second, the Courant algebroid appears in double field theory after imposing a section condition. One might therefore want to formulate a full double gauge theory, related to ours only after the section condition is imposed. Such a double gauge theory might have interesting applications in effectively describing string theory dualities. Third, it remains to be seen whether we can write down six-dimensional maximally superconformal gauge equations which are less restrictive than those obtained in \cite{Lambert:2010wm}, using generalized higher gauge theory. Finally, as stated above, it would be most interesting to extend the twistor descriptions of \cite{Saemann:2012uq,Saemann:2013pca,Jurco:2014mva} to generalized higher gauge theory and to explore the possibility of genuinely non-trivial and non-abelian generalized principal 2-bundles with connection.

\section{Kinematical description}

We begin by reviewing the notion of N$Q$-manifolds and their relation to $L_\infty$-algebras. We use this language to describe ordinary gauge theory in terms of morphisms of graded manifolds and show how this extends to higher gauge theory, following \cite{Atiyah:1957,Bojowald:0406445,Kotov:2007nr,Sati:0801.3480,Fiorenza:2011jr,Gruetzmann:2014ica}. This formulation naturally allows for a generalization to gauge theory involving the exact Courant algebroid $TM\oplus T^*M$.

\subsection{\texorpdfstring{N$Q$}{NQ}-manifolds}
\label{subsec:nq}

Formally, an {\em N-manifold} is a locally ringed space $\CM = \left( M,\CO_\CM\right)$, where $M$ is a manifold and $\CO_\CM$ is an $\NN$-graded commutative ring replacing the ordinary structure sheaf over $M$. More explicitly, we can think of an N-manifold $\CM$ as a tower of fibrations
\begin{equation}
\CM_0 \leftarrow \CM_1 \leftarrow \CM_2 \leftarrow \CM_3\leftarrow \ldots~,
\end{equation}
where $\CM_0=M$ is a manifold and $\CM_i$ for $i\geq 1$ are linear spaces with coordinates of degree $i$, generating the structure sheaf. For more details on this, see e.g.\ \cite{Roytenberg:0203110}. A \textit{morphism of N-manifolds} is then a morphism of graded manifolds $\phi: \left(M,\CO_\CM\right)\to\left(N,\CO_\CN\right)$. In more detail, we have a map $\phi_0: M\to N$ between the underlying manifolds and a degree-preserving map $\phi^\ast: \CO_\CN\to\CO_\CM$ between the structure sheaves, which restricts to the pullback along $\phi_0$ on the sheaf of smooth function on $N$, $\CO_N\subset \CO_\CN$. Note that for higher degrees, $\phi^\ast$ is completely defined by its image on the local coordinates that generate $\CO_\CN$.

An N$Q$-manifold is now an N-manifold $(M,\CO_\CM)$ together with a homological vector field $Q$, that is, a vector field of degree $1$ squaring to zero: $Q^2=0$. The algebra of functions $\CC^\infty(\CM)$ on $\CM$ given by global sections of $\CO_\CM$ together with $Q$ now forms a {\em differential graded algebra}. A \textit{morphism of N$Q$-manifolds} is then a morphism $\phi$ between N$Q$-manifolds $(M,\CO_\CM,Q_\CM)$ and $(N,\CO_\CN,Q_\CN)$ that respects the derivation $Q$, i.e.\ $\phi^\ast \circ Q_\CN = Q_\CM \circ \phi^\ast$.

Physicists may be familiar with N$Q$-manifolds from BRST quantization, where the coordinate degree and $Q$ correspond to the ghost number and the BRST charge, respectively.

A basic example of an N$Q$-manifold is given by $T[1]M$, where we always use $\left[n\right]$ to denote a shift of the degree of some linear space (often the fibers of a vector bundle) by $n$. On $T[1]M$, we have coordinates $\left(x^\mu, \xi^\mu\right)$ on the base and the fibers of degree $0$ and $1$ respectively, i.e.\ we have an N-manifold concentrated in the lowest two degrees. Note that the algebra of functions on $T[1]M$ can be identified with the differential forms $\Omega^\bullet(M)$. Moreover, endowing $T[1]M$ with the homological vector field $Q=\xi^\mu \der{x^\mu}$ promotes it to an N$Q$-manifold. In the identification $\CC^\infty(T[1]M)\cong \Omega^\bullet(M)$, $Q$ becomes the de Rham differential.

A more involved example is $\CV_2\coloneqq T^\ast[2]T[1]M$. Recall that the functor $T^\ast$ gives extra coordinates with opposite degree to the fibers in $T[1]M$. Therefore, local coordinates $\left(x^\mu, \xi^\mu, \xi_\mu, p_\mu\right)$, $\mu=1,\ldots,\dim(M)$, on $T^*[2]T[1]M$ are of degree $0,1,1$ and $2$, respectively. For convenience, we group the coordinates of degree $1$ into a single $\xi^M=(\xi^\mu,\xi_\mu)$, where the index $M$ runs from 1 to $2\dim(M)$. A canonical choice of homological vector field is now
\begin{equation}\label{eq:hom_vec_Courant}
 Q_{\CV_2}=\xi^\mu\der{x^\mu}+p_\mu\der{\xi_\mu}~,
\end{equation}
which can be ``twisted,'' e.g., to 
\begin{equation}\label{eq:hom_vec_Courant_twisted}
 \tilde Q_{\CV_2}=\xi^\mu\der{x^\mu}+p_\mu\der{\xi_\mu}+\frac12 T_{\mu\nu\kappa}\xi^\mu\xi^\nu\der{\xi_\kappa}+\frac{1}{3!}\der{x^\mu}T_{\nu\kappa\lambda}\xi^\nu\xi^\kappa\xi^\lambda\der{p_\mu}~,
\end{equation}
where $T=\tfrac{1}{3!}T_{\mu\nu\kappa}\dd x^\mu\wedge \dd x^\nu\wedge \dd x^\kappa$ is a closed 3-form on $M$, cf.\ \cite{Roytenberg:0203110}. We shall work mostly with the case $T=0$. Altogether, we arrive at an N$Q$-manifold concentrated in degrees $0$ to $2$. This N$Q$-manifold is the one underlying the exact Courant algebroid $TM\oplus T^*M$, and we will come back to this point later. Also, this example is part of a larger class of N$Q$-manifolds given by $\CV_n\coloneqq T^\ast\left[n\right]T[1]M$ containing the \textit{Vinogradov algebroids} $TM\oplus \wedge^{n-1}T^\ast M$. For more details, see e.g.\ \cite{Ritter:2015ffa}.

Another important example of N$Q$-manifolds is that of a grade-shifted Lie algebra $\frg[1]$ with basis $(\tau_\alpha)$ of degree $0$ and coordinates $(w^\alpha)$ of degree $1$. The algebra of functions is given by $\wedge^\bullet\frg^*\cong \sSym(\frg[1]^*)$ and $Q$ is necessarily of the form
\begin{equation}
Q = -\frac12 f^\alpha_{\beta\gamma}w^\beta w^\gamma \der{w^\alpha}~,
\end{equation}
where $f^\alpha_{\beta\gamma}$ are the structure constants of the Lie algebra $\frg$. The condition $Q^2=0$ directly translates to the Jacobi identity. This alternative description of a (finite-dimensional) Lie algebra is the well-known Chevalley-Eilenberg algebra CE$(\frg[1])$ of $\frg$ and we can thus think of a Lie algebra as an N$Q$-manifold concentrated in degree 1. Analogously, we will refer to the differential graded algebra consisting of the algebra of functions on an N$Q$-manifold $\CM$ together with the differential given by the homological vector field as the Chevalley-Eilenberg algebra ${\rm CE}(\CM)$ of $\CM$.

We can readily extend the last example, replacing the shifted Lie algebra by some shifted graded vector space, which we also denote by $\frg[1]$. On the latter, we introduce a basis $\tau_A$ of degree $0$ and coordinates $Z^A$ of degree $\left|A\right|\in\NN$ in $\frg[1]$. The vector field $Q$ is then of the form
\begin{equation}
Q = \sum\limits_{k=1}^\infty \frac{(-1)^{\frac{k(k+1)}{2}}}{k!}Z^{B_1}\cdots Z^{B_k}m^A_{B_1\ldots B_k} \der{Z^A}~,
\end{equation}
where $m^A_{B_1\ldots B_k}$ can be non-zero only if $\sum_{i=1}^k \left|B_i\right| = \left|A\right|+1$ since $Q$ is of degree~$1$. The minus signs and normalizations are chosen for convenience. 

We now also introduce a basis $\hat{\tau}_A$ on the unshifted $\frg$, where we absorb all grading in the basis instead of the coordinates. Thus, $\hat{\tau}_A$ has degree $\left|A\right|-1$. The structure constants $m^A_{ B_1\ldots  B_k}$ can then be used to define the following graded antisymmetric, $k$-ary brackets $\mu_k$ on $\frg$ of degree $k-2$:
\begin{equation}\label{eq:def_homotopy_products}
\mu_k(\hat{\tau}_{ B_1},\ldots,\hat{\tau}_{ B_k}) = m^A_{ B_1\ldots  B_k} \hat{\tau}_A~.
\end{equation}
For an N$Q$-manifold concentrated in degrees $1$ to $n$, the condition $Q^2=0$ amounts to the homotopy Jacobi relations of an $n$-term $L_\infty$-algebra with higher products $\mu_k$, cf.\ \cite{Lada:1992wc,Lada:1994mn}. Such $n$-term $L_\infty$-algebras are expected to be categorically equivalent to semistrict Lie $n$-algebras.

As a constructive example let us look at a $2$-term $L_\infty$-algebra originating from the N$Q$-manifold $W[1]\leftarrow V[2]$. In a basis $(\tau_\alpha)$ and $(t_a)$ with corresponding coordinates $w^\alpha$ and $v^a$ of degree $1$ and $2$ on $W[1]$ and $V[2]$, respectively, the vector field $Q$ reads as
\begin{equation}
\label{eq:lie2q}
Q = - m^\alpha_a v^a \der{w^\alpha} - \frac12 m^\alpha_{\beta\gamma} w^\beta w^\gamma \der{w^\alpha} 
- m^a_{\alpha b} w^\alpha v^b \der{v^a} + \frac{1}{3!} m^a_{\alpha\beta\gamma} w^\alpha
w^\beta w^\gamma \der{v^a}~.
\end{equation}
We define corresponding $L_\infty$-algebra products $\mu_1$, $\mu_2$ and $\mu_3$ on $W\leftarrow V[1]$ via \eqref{eq:def_homotopy_products} and the condition $Q^2=0$ leads to higher homotopy relations, which, in terms of the graded basis $(\hat{\tau}_\alpha)$ and $(\hat{t}_a)$ on $W\leftarrow V[1]$, are
\begin{subequations}
\begin{align}
\mu_1(\mu_1(\hat{t}_a)) &= 0~, \nonumber\\
\mu_1(\mu_2(\hat{\tau}_\alpha,\hat{t}_a)) &= \mu_2(\hat{\tau}_\alpha,\mu_1(\hat{t}_a))~, \nonumber\\
\mu_2(\mu_1(\hat{t}_a),\hat{t}_b) &= \mu_2 (\hat{t}_a, \mu_1(\hat{t}_b))~,\\
\mu_1(\mu_3(\hat{\tau}_\alpha,\hat{\tau}_\beta,\hat{\tau}_\gamma)) &=
-\mu_2(\mu_2(\hat{\tau}_\alpha,\hat{\tau}_\beta),\hat{\tau}_\gamma)
-\mu_2(\mu_2(\hat{\tau}_\gamma,\hat{\tau}_\alpha),\hat{\tau}_\beta)
-\mu_2(\mu_2(\hat{\tau}_\beta,\hat{\tau}_\gamma),\hat{\tau}_\alpha)~, \nonumber\\
\mu_3(\mu_1(\hat{t}_a),\hat{\tau}_\alpha,\hat{\tau}_\beta) &=
-\mu_2(\mu_2(\hat{\tau}_\alpha,\hat{\tau}_\beta),\hat{t}_a)
-\mu_2(\mu_2(\hat{t}_a,\hat{\tau}_\alpha),\hat{\tau}_\beta)
-\mu_2(\mu_2(\hat{\tau}_\beta,\hat{t}_a),\hat{\tau}_\alpha)~,\nonumber
\end{align}
\text{and}
\begin{align}
0&= \mu_2(\mu_3(\hat{\tau}_\alpha,\hat{\tau}_\beta,\hat{\tau}_\gamma),\hat{\tau}_\delta)
-\mu_2(\mu_3(\hat{\tau}_\delta,\hat{\tau}_\alpha,\hat{\tau}_\beta),\hat{\tau}_\gamma) 
+\mu_2(\mu_3(\hat{\tau}_\gamma,\hat{\tau}_\delta,\hat{\tau}_\alpha),\hat{\tau}_\beta)\nonumber\\
&\phantom{=}
-\mu_2(\mu_3(\hat{\tau}_\beta,\hat{\tau}_\gamma,\hat{\tau}_\delta),\hat{\tau}_\alpha)
-\mu_3(\mu_2(\hat{\tau}_\alpha,\hat{\tau}_\beta),\hat{\tau}_\gamma,\hat{\tau}_\delta)
+\mu_3(\mu_2(\hat{\tau}_\beta,\hat{\tau}_\gamma),\hat{\tau}_\delta,\hat{\tau}_\alpha)\nonumber\\
&\phantom{=}
-\mu_3(\mu_2(\hat{\tau}_\gamma,\hat{\tau}_\delta),\hat{\tau}_\alpha,\hat{\tau}_\beta)
+\mu_3(\mu_2(\hat{\tau}_\delta,\hat{\tau}_\alpha),\hat{\tau}_\beta,\hat{\tau}_\gamma)
+\mu_3(\mu_2(\hat{\tau}_\alpha,\hat{\tau}_\gamma),\hat{\tau}_\beta,\hat{\tau}_\delta)\nonumber\\
&\phantom{=}
-\mu_3(\mu_2(\hat{\tau}_\beta,\hat{\tau}_\gamma),\hat{\tau}_\alpha,\hat{\tau}_\delta)~.
\end{align}
\end{subequations}

More generally, if the N$Q$-algebra is concentrated in degrees $0$ to $n$, one analogously obtains an $L_\infty$-algebroid. In fact, $T[1]M$ and $T^\ast[2]T[1]M$ were particular examples of such $L_\infty$-algebroids.

The natural notion of inner product on an $L_\infty$-algebra arises from an additional symplectic structure on the underlying N$Q$-manifold. A {\em symplectic N$Q$-manifold of degree $n$} is an N$Q$-manifold $\CM=(M,\CO_\CM,Q,\omega)$ endowed with a closed, non-degenerate 2-form $\omega$ of degree\footnote{This is the degree in $\CC^\infty(\CM)$, not the form degree of $\omega$.} $n$ satisfying $\CL_Q \omega = 0$. If the degree of $\omega$ is odd, such symplectic N$Q$-manifolds are also known as $Q$P-manifolds \cite{Alexandrov:1995kv} or P-manifolds \cite{Schwarz:1992gs}. In the general case, symplectic N$Q$-manifolds of degree $n$ are also called $\Sigma_n$-manifolds \cite{Severa:2001aa}.

A simple example of a symplectic N$Q$-manifold of degree $1$ is $T^\ast[1]M$ with coordinates $\left(x^\mu,p_\mu\right)$, homological vector field $Q=\pi^{\mu\nu} p_\mu \der{x^\nu}$ for some anti-symmetric bivector $\pi^{\mu\nu}$ and symplectic form $\omega = \dd x^\mu \wedge \dd p_\mu$. Indeed, $\CL_Q\omega=\dd\iota_Q \omega = \pi^{\mu\nu}\dd p_\mu\wedge\dd p_\nu = 0$.

We are mostly interested in the N$Q$-manifold $\CV_2\coloneqq T^*[2]T[1]M$ with coordinates $\left(x^\mu,\xi^\mu,\xi_\mu,p_\mu\right) = \left(x^\mu,\xi^M,p_\mu\right)$ as defined above. With
\begin{equation}\label{eq:symp_struct_Courant}
\omega = \dd x^\mu \wedge \dd p_\mu + \dd \xi^\mu \wedge \dd \xi_\mu~, 
\end{equation}
$\CV_2$ becomes a symplectic N$Q$-manifold of degree 2: we have
\begin{equation}
\CL_{Q_{\CV_2}} \omega = \dd\iota_{Q_{\CV_2}} \omega = \dd \xi^\mu \wedge \dd p_\mu + \dd p_\mu \wedge \dd \xi^\mu = 0~,
\end{equation}
where $Q_{\CV_2}$ is the homological vector field \eqref{eq:hom_vec_Courant}. The symplectic structure \eqref{eq:symp_struct_Courant} is also compatible with the twisted homological vector field \eqref{eq:hom_vec_Courant_twisted}.

As shown in \cite{Roytenberg:0203110}, the data specifying a symplectic N$Q$-structure on $T^*[2]T[1]M$ are equivalent to the data specifying a Courant algebroid structure on the bundle $TM\oplus T^*M$. In particular, sections of the Courant algebroid are functions in $\CC^\infty(T^*[2]T[1]M)$ which are linear in the coordinates $\xi^M$. Moreover, a metric on sections of $TM\oplus T^*M$ originates from the Poisson bracket induced by the symplectic structure \eqref{eq:symp_struct_Courant}. With the symplectic structure $\omega=\tfrac12 \dd Z^A\wedge \omega_{AB}\dd Z^B$ in coordinates $Z^A=(x^\mu,\xi^M,p_\mu)$, we have
\begin{equation}
 (a_M\xi^M,b_N\xi^N):=\tfrac12\{a_M\xi^M,b_N\xi^N\}=\frac12\left(a_M\xi^M\right)\frac{\overleftarrow\partial
    }{\partial Z^A}\omega^{AB}\frac{\overrightarrow\partial
    }{\partial Z^B}\left(b_N\xi^N\right)~,
\end{equation}
where $\omega^{AB}$ is the inverse matrix to $\omega_{AB}$. With our choice of symplectic structure \eqref{eq:symp_struct_Courant}, we have
\begin{equation}
  (a_M\xi^M,b_N\xi^N)=\tfrac12\left(\iota_{a^\mu\der{x^\mu}}b_\nu\dd x^\nu+\iota_{b^\nu\der{x^\nu}}a_\mu\dd x^\mu\right)=\tfrac12(a^\mu b_\mu+b^\mu a_\mu)~.
\end{equation}
For simplicity, we will refer to both the symplectic N$Q$-manifolds $\CV_2=T^*[2]T[1]M$ and the vector bundle $TM\oplus T^*M$ with Courant algebroid structure as Courant algebroid. 

Note that the exact Courant algebroid $TM\oplus T^*M$ features prominently in generalized geometry and double field theory. We therefore expect our following constructions to be relevant in this context.

\subsection{Gauge connections as morphisms of N-manifolds}\label{ssec:ordinary_gauge}

In ordinary gauge theory, we consider connections on principal $\sG$-bundles over some manifold $M$ and encode them locally as 1-forms $A$ taking values in a Lie algebra $\frg=\sLie(\sG)$. The curvature of $A$ is $F\coloneqq \dd A +\tfrac12\left[A,A\right]$ and gauge transformations are parameterized by $\sG$-valued functions $g$ and act on $A$ via $A\mapsto\tilde{A} = g^{-1}Ag+g^{-1}\dd g$. At infinitesimal level, these are given by $A\mapsto A+\delta A$, where $\delta A = \dd \lambda + \left[A, \lambda\right]$ with $\lambda $ a $\frg$-valued function.

Let us now reformulate the local description of gauge theory using morphisms of N-manifolds. As discussed in section \ref{subsec:nq}, differential forms can be encoded as functions on the N$Q$-manifold $T[1]M$. In terms of coordinates $\left(x^\mu,\xi^\mu\right)$ of degree $0$ and $1$, respectively, the de Rham differential corresponds to the homological vector field $Q_{T[1]M}=\xi^\mu\der{x^\mu}$. We also regard $\frg$ as an N$Q$-manifold $\frg[1]$ with coordinates $w^\alpha$ of degree $1$ and $Q_\frg=-\tfrac12f^\alpha_{\beta\gamma}w^\beta w^\gamma\der{w^\alpha}$, cf.\ again section \ref{subsec:nq}. 

A local connection 1-form $A$ is then encoded in a morphism of N-manifolds $a$ between these two N$Q$-manifolds
\begin{equation*}
T[1]M\overset{a}{\longrightarrow}\frg[1]~.
\end{equation*}
Recall that it suffices to define the action of $a$ on the local coordinates of $\frg[1]$, so we define
\begin{equation}
 A=A^\alpha\hat\tau_\alpha ~~~\mbox{with}~~~A^\alpha = A^\alpha_\mu \xi^\mu\coloneqq a^\ast(w^\alpha)~,
\end{equation}
where $(\hat \tau_\alpha)$ is a basis on $\frg$. The curvature $F$ of $A$ then describes the failure of $a$ to be a morphism of N$Q$-manifolds:
\begin{equation}
 F=F^\alpha\hat\tau_\alpha ~~~\mbox{with}~~~F^\alpha = \left(Q_{T[1]M}\circ a^\ast-a^\ast\circ Q_\frg\right) (w^\alpha)~.
\end{equation}
Indeed, we have 
\begin{equation}
\begin{aligned}
F^\alpha&=
Q_{T[1]M} A^\alpha - a^\ast\left(-\tfrac12f^\alpha_{\beta\gamma}w^\beta w^\gamma\right)\\
&= \left(\xi^\mu\der{x^\mu} A\right)^\alpha + \frac12\mu_2(A,A)^\alpha~,
\end{aligned}
\end{equation}
where $\mu_2$ denotes the Lie bracket on $\frg$. 

Gauge transformations between $a$ and $\tilde{a}$ are encoded in flat homotopies between these, that is, morphisms $T[1](M\times [0,1])\stackrel{\hat{a}}{\longrightarrow}\frg[1]$ which are flat along the additional direction \cite{Fiorenza:2010mh}. More precisely, given coordinates $r$ along $[0,1]$ and $\rho$ on $T[1][0,1]$, we have
\begin{equation}
 \hat{a}|_{r=0}=a\eand \hat{a}|_{r=1}=\tilde{a}~.
\end{equation}
Note that $\hat{a}$ defines a gauge potential 
\begin{equation}
\hat{a}^*(w^\alpha)=\hat{A}^\alpha\coloneqq\hat A^\alpha_\mu(x,r)\xi^\mu+\hat A^\alpha_r(x,r)\rho
\end{equation}
and a curvature
\begin{equation}
\hat F=\hat F_{\mu\nu}\xi^\mu\xi^\nu+\underbrace{\left(\der{x^\mu}\hat A_r(x,r)+\mu_2(\hat A_\mu(x,r),\hat A_r(x,r))-\der{r}\hat A_\mu(x,r)\right)\xi^\mu r}_{\hat{F}_\perp}~,
\end{equation}
where we used the amended $\hat Q_{T[1]M}= \xi^\mu\der{x^\mu} + \rho \der{r}$ on $T[1](M\times \left[0,1\right])$. For the homotopy $\hat a$ to be flat, we require $\CF_{\perp}=0$, which implies that
\begin{equation}\label{eq:inft_gauge}
 \der{r}\hat A_\mu(x,r)=\der{x^\mu}\hat A_r(x,r)+\mu_2(\hat A_\mu(x,r),\hat A_r(x,r))~.
\end{equation}
Restricting to $r=0$ yields the usual formula for gauge transformations with infinitesimal gauge parameter $\lambda =A_r(x,0)$. Integrating \eqref{eq:inft_gauge} with the boundary condition $\hat A_\mu(x,0)=A_\mu(x)$, we obtain the finite form 
\begin{equation}
 A(x,1)=g^{-1}(x) A(x,0) g(x)+g^{-1}(x)\dd g(x)~, 
\end{equation}
where $g(x)$ is the path-ordered exponential of $A_r(x,r)$ along $[0,1]$.

\subsection{Higher gauge connections}

We can readily extend the picture of the previous section to the case of higher gauge theory. Here, we simply replace the gauge Lie algebra by a general $L_\infty$-algebra $\frg$.\footnote{In principle, we could also allow for $L_\infty$-algebroids, which would lead us to higher gauged sigma models.} The morphism of N-manifolds $a:T[1]M\rightarrow \frg[1]$ now also contains forms of higher degree. Similarly, the curvature, which is again given by the failure of $a$ to be a morphism of N$Q$-manifolds, leads to higher curvature forms.

As an instructive example let us look at the 2-term $L_\infty$-algebra $W\leftarrow V[1]$ introduced before in section \ref{subsec:nq}. The image of the pullback morphism $a^*$ on the coordinates $(w^\alpha)$ and $(v^a)$ of degree $1$ and $2$ on the shifted vector space $W[1]\leftarrow V[2]$ is given by 
\begin{equation}
a^\ast (w^\alpha) = A^\alpha_\mu \xi^\mu \eand a^\ast (v^a) = \tfrac12B^a_{\mu\nu} \xi^\mu\xi^\nu~, 
\end{equation}
where in addition to the $W$-valued $1$-form potential $A$ we now also have a $V[1]$-valued $2$-form potential $B$. We combine both into the 2-connection
\begin{equation}
 \CA=A_\mu\xi^\mu+\tfrac12 B_{\mu\nu}\xi^\mu\xi^\nu~.
\end{equation}
With $Q_\frg$ from \eqref{eq:lie2q}, we compute the curvature components
\begin{equation}
 F^\alpha = \left(Q_{T[1]M}\circ a^\ast-a^\ast\circ Q_\frg\right) (w^\alpha)~,~~~H^a=\left(Q_{T[1]M}\circ a^\ast-a^\ast\circ Q_\frg\right) (v^a)
\end{equation}
to be 
\begin{equation}
\begin{aligned}
F^\alpha\hat\tau_\alpha&=\tfrac12 F_{\mu\nu}\xi^\mu\xi^\nu=\left(\der{x^\mu} A_\nu + \frac12\mu_1(B_{\mu\nu}) +\frac12 \mu_2(A_\mu,A_\nu)\right)\xi^\mu\xi^\nu~,\\
H^a t_a&=\tfrac{1}{3!}H_{\mu\nu\kappa}\xi^\mu\xi^\nu\xi^\kappa= \left(\frac12 \der{x^\mu} B_{\nu\kappa} + \frac12 \mu_2(A_\mu,B_{\nu\kappa}) - \frac{1}{3!}\mu_3(A_\mu,A_\nu,A_\kappa)\right)\xi^\mu\xi^\nu\xi^\kappa~,
\end{aligned}
\end{equation}
which we combine into the 2-curvature
\begin{equation}
\CF \eqqcolon \tfrac12 F_{\mu\nu}\xi^\mu\xi^\nu + \tfrac{1}{3!}H_{\mu\nu\kappa}\xi^\mu \xi^\nu \xi^\kappa~.
\end{equation}

Again, the infinitesimal gauge transformations between $a$ and $\tilde{a}$ are encoded in homotopies $\hat{a}: T[1](M\times [0,1]) \to \frg[1]$ that are flat in the extra homotopy direction. We use coordinates $(x,\xi,r,\rho)$ on $T[1](M\times [0,1])$ and we have $\left.\hat{a}\right|_{r=0} = a$ as well as $\left.\hat{a}\right|_{r=1}=\tilde{a}$. Then $\hat{a}$ defines a gauge potentials as before, that is,
\begin{equation}
\begin{aligned}
\hat{a}^\ast (w^\alpha) &= \hat{A}^\alpha \coloneqq \hat{A}^\alpha_\mu \xi^\mu + \hat{A}^\alpha_r \rho~, \\
\hat{a}^\ast (v^a) &= \hat{B}^a \coloneqq \tfrac12\hat{B}^a_{\mu\nu} \xi^\mu\xi^\nu + \hat{B}^a_{\mu r} \xi^\mu \rho~.
\end{aligned}
\end{equation}
Using the extended vector field $\hat Q_{T[1]M}=\xi^\mu \der{x^\mu} + \rho \der{r}$, we calculate the curvature defined by $\hat{a}$ along the additional direction to be
\begin{equation}
\begin{aligned}
\hat{F}_{\mu r}\xi^\mu\rho &= \left(\der{x^\mu} \hat{A}_r - \der{r}\hat{A}_\mu + \mu_2(\hat{A}_\mu,\hat{A}_r)+\mu_1(\hat{B}_{\mu r})\right) \xi^\mu \rho~,\\
\tfrac12\hat{H}_{\mu\nu r} \xi^\mu\xi^\nu\rho&= \left( \frac12 \der{r} \hat{B}_{\mu\nu} + \der{x^\mu} \hat{B}_{\nu r} +\frac12\mu_2(\hat{A}_r, \hat{B}_{\mu\nu})\right.\\
&\phantom{= \hat{\CF}_\parallel +}\left.\phantom{\der{r} \hat{B}_{\mu\nu} + \der{x^\mu} \hat{B}_{\nu r}}+\mu_2(\hat{A}_\mu,\hat{B}_{\nu r}) -\frac12\mu_3(\hat{A}_\mu,\hat{A}_\nu,\hat{A}_r)\right) \xi^\mu\xi^\nu\rho~.
\end{aligned}
\end{equation}
As before, the infinitesimal gauge transformations are encoded in the flat homotopies for which the above curvature in the directions including $\rho$ vanishes. This leads to the transformations
\begin{equation}
\begin{aligned}
\der{r} \hat{A}_\mu &= \der{x^\mu} \hat{A}_r + \mu_2(\hat{A}_\mu,\hat A_r) + \mu_1(\hat{B}_{\mu r})~,\\
\der{r} \hat{B}_{\mu\nu} &= -2\der{x^\mu} \hat{B}_{\nu r} + \mu_2(\hat A_r,\hat{B}_{\nu\mu}) - 2\mu_2(\hat A_\mu,\hat{B}_{\nu r}) +\mu_3(\hat A_\mu,\hat A_\nu,\hat A_r)~,
\end{aligned}
\end{equation}
which are parameterized by two infinitesimal gauge parameters: the $W$-valued function $\hat A_r(x,0)$ and a $V[1]$-valued $1$-form $\hat{B}_{\nu r}(x,0)$. We thus obtain the infinitesimal gauge transformations of semistrict higher gauge theory as found e.g.\ in \cite{Jurco:2014mva}.\footnote{An alternative approach to finite gauge transformations of semistrict higher gauge theory is found in \cite{Zucchini:2011aa}.} Putting $\mu_3=0$, we obtain the infinitesimal gauge transformations of strict higher gauge theory, which can be integrated as done in \cite{Demessie:2014ewa}. Setting $\hat{B}, \mu_1$ and $\mu_3$ to zero reduces the transformation back to the case of ordinary gauge theory.

\subsection{Local description of generalized higher gauge theory}
\label{subsec:localdblhg}

We now come to our extension of higher gauge theory to {\em generalized higher gauge theory}. To this end, we replace the domain of the morphism of N-manifolds, which has been $T[1]M$ so far, by the Courant algebroid $\CV_2=T^*[2]T[1]M$ with coordinates $\left(x^\mu, \xi^M, p_\mu\right)$ of degree $0,1$ and $2$, respectively, and homological vector field $Q_{\CV_2} = \xi^\mu \der{x^\mu} + p_\mu \der{\xi_\mu}$, see section \ref{subsec:nq}.

Generalized higher gauge theory is thus given by a morphism of N-manifolds $a:\CV_2 \to \frg[1]$, where $\frg$ is an arbitrary $L_\infty$-algebra. We again focus on the example where $\frg$ is a $2$-term $L_\infty$-algebra $W\leftarrow V[1]$ and we introduce a basis $(\tau_\alpha,t_a)$ and coordinates $(w^\alpha,v^a)$ of degree $1$ and $2$, respectively, on $\frg[1]$. The homological vector field $Q_\frg$ is given in (\ref{eq:lie2q}). The images of the coordinates of $\frg[1]$ under the morphism $a$ are
\begin{equation}
\begin{aligned}
a^\ast (w^\alpha) &= A^\alpha_M \xi^M~,\\
a^\ast (v^a) &= \tfrac12 B^a_{MN}\xi^M\xi^N + B^{a\mu}p_\mu~,
\end{aligned}
\end{equation}
where $A=A_M\xi^M=A_\mu\xi^\mu+A^\mu \xi_\mu$ can now be regarded as the sum of a 1-form and a vector field, which are both $W$-valued. Similarly, $B$ consists of a 2-form, a bivector, a tensor of degree (1,1) and a vector field, all taking values in $V[1]$. We combine all these into the generalized 2-connection
\begin{equation}\label{eq:gen_2_connection}
 \CA=A_\mu\xi^\mu+A^\mu \xi_\mu+\tfrac12 B_{\mu\nu}\xi^\mu\xi^\nu+B_{\mu}{}^\nu\xi^\mu\xi_\nu+\tfrac12 B^{\mu\nu}\xi_\mu\xi_\nu + B^{\mu}p_\mu~.
\end{equation}
The generalized 2-curvature $\CF$ is again obtained from the failure of $a$ to be a morphism of N$Q$-manifolds, and splits into components according to
\begin{equation}\label{eq:gen_2_curvature_components}
\begin{aligned}
\CF &= \tfrac12 F_{\mu\nu}\xi^\mu\xi^\nu + F_{\mu}^{\phantom{\mu}\nu}\xi^\mu\xi_\nu + \tfrac12F^{\mu\nu}\xi_\mu\xi_\nu + F^\mu p_\mu\\
&\phantom{=} +\tfrac{1}{3!}H_{\mu\nu\kappa}\xi^\mu\xi^\nu\xi^\kappa + \tfrac{1}{2}H_{\mu\nu}^{\phantom{\mu\nu}\kappa} \xi^\mu\xi^\nu\xi_\kappa + \tfrac{1}{2}H_{\mu}^{\phantom{\mu}\nu\kappa} \xi^\mu \xi_\nu\xi_\kappa\\
&\phantom{=} + \tfrac{1}{3!}H^{\mu\nu\kappa}\xi_\mu\xi_\nu\xi_\kappa+
H_{\mu}^{\phantom{\mu}\nu}\xi^\mu p_\nu + H^{\mu\nu} \xi_\mu p_\nu~.
\end{aligned}
\end{equation}
The components of $\CF$ are computed to be 
\begin{equation}\label{eq:generalized_2_curvature}
 \begin{aligned}
  \CF= &\left(\der{x^\mu} A_\nu + \frac12 \mu_2 (A_\mu,A_\nu) + \frac12\mu_1 (B_{\mu\nu})\right)\xi^\mu\xi^\nu+\\
  &+\left(\der{x^\mu} A^\nu + \mu_2(A_\mu,A^\nu) + \mu_1(B_\mu^{\phantom{\mu}\nu})\right)\xi^\mu\xi_\nu+\\
  &+\left(\frac12 \mu_2(A^\mu,A^\nu) + \frac12\mu_1(B^{\mu\nu})\right)\xi_\mu\xi_\nu+\left(A^\mu + \mu_1 (B^\mu)\right)p_\mu+\\
  &+\left(-\frac{1}{3!}\mu_3(A_\mu,A_\nu,A_\kappa)+ \frac12\mu_2(A_\mu,B_{\nu\kappa}) +\frac12\der{x^\mu} B_{\nu\kappa}\right)\xi^\mu\xi^\nu\xi^\kappa+\\
  &+\left(-\frac12\mu_3(A_\mu,A_\nu,A^\kappa)+ \mu_2(A_\mu,B_{\nu}^{\phantom{\nu}\kappa}) + \frac12\mu_2(A^\kappa,B_{\mu\nu})+ \der{x^\mu} B_{\nu}^{\phantom{\nu}\kappa}\right)\xi^\mu\xi^\nu\xi_\kappa+\\
  &+\left(-\frac12\mu_3(A_\mu,A^\nu,A^\kappa)+ \frac12\mu_2(A_\mu,B^{\nu\kappa}) - \mu_2(A^\nu,B_\mu^{\phantom{\mu}\kappa})+ \frac12\der{x^\mu} B^{\nu\kappa}\right)\xi^\mu\xi_\nu\xi_\kappa+\\
  &+\left(-\frac{1}{3!}\mu_3(A^\mu,A^\nu,A^\kappa)+ \frac12\mu_2(A^\mu,B^{\nu\kappa})\right)\xi_\mu\xi_\nu\xi_\kappa+\\
  &+\left(\mu_2(A_\mu,B^\nu) + B_\mu^{\phantom{\mu}\nu} + \der{x^\mu} B^\nu\right)\xi^\mu p_\nu+\left(\mu_2(A^\mu,B^\nu)+B^{\mu\nu}\right)\xi_\mu p_\nu~.
 \end{aligned}
\end{equation}

Flat homotopies between morphism of N-manifolds $a$ and $\tilde{a}$ give the generalized higher gauge transformations. These are encoded in morphisms $\hat a$ from $T^\ast[2]T[1]M \times T[1]I$ to $\frg[1]$, where we introduce additional coordinates $(r,\rho)$ of degrees $(0,1)$ in the new direction and the vector field $Q_{\CV_2}$ is amended to
\begin{equation}
\hat Q_{\CV_2} = \xi^\mu \der{x^\mu} + p_\mu \der{\xi_\mu} + \rho \der{r}~.
\end{equation}
The morphism $\hat a$ then defines gauge potentials
\begin{align}
\hat a^\ast(w^\alpha) &= \hat A^\alpha \coloneqq \hat{A}^\alpha_M \xi^M + \hat{A}^\alpha_r \rho~, \\
\hat a^\ast(v^a) &= \hat B^a \coloneqq \tfrac12\hat{B}^a_{MN} \xi^M\xi^N + \hat{B}^a_{M r} \xi^M\rho + \hat{B}^{a\mu}p_\mu~,
\end{align}
which lead to curvature terms along the homotopy direction
\begin{equation*}
 \begin{aligned}
  \CF_\perp=&\left(-\der{r} \hat{A}_\mu  + \der{x^\mu} \hat{A}_r +\mu_1(\hat{B}_{\mu r}) + \mu_2(\hat{A}_\mu,\hat{A}_r)\right)\xi^\mu\rho+\\
&+\left(-\der{r} \hat{A}^\mu + \mu_1(\hat{B}^{\mu}_{\phantom{\mu} r}) + \mu_2(\hat{A}^\mu, \hat{A}_r)\right)\xi_\mu \rho+\\
&+\left(\frac12 \der{r} \hat{B}_{\mu\nu} +\der{x^\mu} \hat{B}_{\nu r} + \mu_2(\hat{A}_\mu,\hat{B}_{\nu r}) + \frac12 \mu_2(\hat{A}_r,\hat{B}_{\mu\nu}) -\frac12 \mu_3(\hat{A}_\mu,\hat{A}_\nu,\hat{A}_r)\right)\xi^\mu\xi^\nu\rho+\\
&+\left(\der{r} \hat{B}_{\mu}^{\phantom{\mu}\nu} +\der{x^\mu} \hat{B}^{\nu}_{\phantom{\nu} r} + \mu_2(\hat{A}_\mu,\hat{B}^{\nu}_{\phantom{\nu} r}) - \mu_2(\hat{A}^\nu,\hat{B}_{\mu r})\right. \\
&\hspace{3cm}\left.\phantom{\der{r}}+\mu_2(\hat{A}_r,\hat{B}_\mu^{\phantom{\mu}\nu}) - \mu_3(\hat{A}_\mu,\hat{A}^\nu,\hat{A}_r)\right)\xi^\mu\xi_\nu \rho+\\
&+\left(\frac12\der{r} \hat{B}^{\mu\nu} +\mu_2(\hat{A}^\mu,\hat{B}^{\nu}_{\phantom{\nu}r}) +\frac12\mu_2(\hat{A}_r,\hat{B}^{\mu\nu}) - \frac12\mu_3(\hat{A}^\mu,\hat{A}^\nu, \hat{A}_r)\right)\xi_\mu\xi_\nu\rho+\\
&+\left(B^{\mu}_{\phantom{\mu} r}-\der{r} \hat{B}^\mu -\mu_2(\hat{A}_r,\hat{B}^\mu)\right)p_\mu\rho~.
 \end{aligned}
\end{equation*}
The requirement that these terms vanish yields the infinitesimal gauge transformations
\begin{equation}\label{eq:infinit_gen_gauge_trafo}
\begin{aligned}
\der{r} \hat{A}_\mu &= \der{x^\mu} \hat{A}_r + \mu_1(\hat{B}_{\mu r}) + \mu_2(\hat{A}_\mu,\hat{A}_r),\\
\der{r} \hat{A}^\mu &= \mu_1(\hat{B}^{\mu}_{\phantom{\mu} r}) + \mu_2(\hat{A}^\mu,\hat{A}_r)~,\\
\der{r} \hat{B}_{\mu\nu} &= -2\der{x^\mu} \hat{B}_{\nu r} - 2\mu_2(\hat{A}_\mu,\hat{B}_{\nu r})-\mu_2(\hat{A}_r,\hat{B}_{\mu\nu})+\mu_3(\hat{A}_\mu,\hat{A}_\nu,\hat{A}_r)~,\\
\der{r} \hat{B}_{\mu}^{\phantom{\mu}\nu} &= -\der{x^\mu} \hat{B}^\nu{}_r - \mu_2(\hat{A}_\mu,\hat{B}^\nu{}_r) + \mu_2(\hat{A}^\nu,\hat{B}_{\mu r}) - \mu_2(\hat{A}_r,\hat{B}_{\mu}^{\phantom{\mu}\nu})  + \mu_3(\hat{A}_\mu,\hat{A}^\nu,\hat{A}_r)~,\\
\der{r} \hat{B}^{\mu\nu} &= -2\mu_2(\hat{A}^\mu,\hat{B}^\nu{}_r) -\mu_2(\hat{A}_r,\hat{B}^{\mu\nu}) + \mu_3(\hat{A}^\mu,\hat{A}^\nu,\hat{A}_r)~,\\
\der{r} \hat{B}^\mu &= B^\mu{}_{ r}-\mu_2(\hat{A}_r,\hat{B}^\mu)~,
\end{aligned}
\end{equation}
which are parameterized by a $W$-valued function $\hat{A}_r$, as well as a $1$-form $\hat{B}_{\mu r}$ and a vector field  $\hat{B}^{\mu}_{\phantom{\mu} r}$, both taking values in $V[1]$.

Note that generalized higher gauge theory contains higher gauge theory. In particular, if we put the fields $A^\mu$, $B_\mu{}^\nu$, $B^{\mu\nu}$ and $B^\mu$ to zero, we obtain the usual 2-connection. Analogously, we can restrict the gauge transformations.

\subsection{Equivalent description from Maurer-Cartan equations}

The above gauge potentials, field strengths and gauge transformations can also be derived in a different manner as we explain now, following a similar discussion as that in \cite{Jurco:2014mva}. First, note that functions $\CC^\infty(\CV_2)$ on the Courant algebra $\CV_2$ form a differential graded algebra with differential $Q_{\CV_2}$. It is well known that the tensor product of such a differential graded algebra with an $L_\infty$-algebra carries a natural $L_\infty$-algebra structure. For us, the relevant $L_\infty$-algebra  is $\frg=(W\leftarrow V[1])$ and the tensor product $\tilde \sL=\CC^\infty(\CV_2)\otimes \frg$ carries the higher products
\begin{equation}
\tilde\mu_i(f_1\otimes\ell_1,\ldots,f_i\otimes \ell_i)\ =\ 
\left\{\begin{array}{ll}
(Q_{\CV_2} f_1) \otimes \ell_1+(-1)^{\deg(f_1)}f_1\otimes \mu_1(\ell_1)&\efor i=1~,\\
\chi~(f_1\cdots f_i)\otimes \mu_i(\ell_1,\ldots,\ell_i)&\efor i=2,3~,\end{array}\right.
\end{equation}
where $\mu_i$ are the higher products in $\frg$, $f_i\in \CC^\infty(\CV_2)$ and $\ell_i\in \frg$, $\deg$ denotes the degree and $\chi=\pm1$ is the Koszul sign arising from moving functions on $\CV_2$ past elements of $\frg$. Note that the total degree of an element $f\otimes \ell$ in $\CC^\infty(\CV_2)\otimes \frg$ is $\deg(f)-\deg(\ell)$ and we truncate $\CC^\infty(\CV_2)\otimes \frg$ to non-negative degrees.

Recall that an element $\phi$ of an $L_\infty$-algebra $\tilde \sL$ is called a {\em Maurer-Cartan element}, if it satisfies the  {\em homotopy Maurer--Cartan equation}
\begin{equation}\label{eq:MCeqs}
 \sum_{i=1}^\infty \frac{(-1)^{i(i+1)/2}}{i!}\tilde \mu_i(\phi,\ldots,\phi)\ =\ 0~.
\end{equation}
This equations is invariant under infinitesimal gauge symmetries parameterized by an element $\lambda\in \tilde \sL$ of degree 0 according to
\begin{equation}\label{eq:MCgaugetrafos}
 \phi\rightarrow \phi+\delta \phi\ewith \delta \phi\ =\ \sum_i \frac{(-1)^{i(i-1)/2}}{(i-1)!}\tilde \mu_i(\lambda,\phi,\ldots,\phi)~,
\end{equation}
cf.\ \cite{Zwiebach:1992ie,Lada:1992wc,Jurco:2014mva}. Equation \eqref{eq:MCeqs} states that the higher curvature vanishes and therefore, it can be used to identify the correct notion of curvature. Equation \eqref{eq:MCgaugetrafos} then gives the appropriate infinitesimal gauge transformations.

Following \cite{Jurco:2014mva}, we now consider an element $\phi$ of degree $1$ in $\sL=\CC^\infty(\CV_2)\otimes \frg$ with $\frg=W\leftarrow V[1]$, which we can identify with the generalized 2-connection of equation \eqref{eq:gen_2_connection}. That is, we write
\begin{equation}
\begin{aligned}
 \phi &=A^\alpha_\mu\xi^\mu\otimes \hat\tau_\alpha+A^{\mu\alpha} \xi_\mu\otimes \hat\tau_\alpha +\tfrac12 B^a_{\mu\nu}\xi^\mu\xi^\nu\otimes \hat t_a \\
&+B_{\mu}{}^{\nu a}\xi^\mu\xi_\nu\otimes \hat t_a+\tfrac12 B^{\mu\nu a}\xi_\mu\xi_\nu\otimes \hat t_a + B^{\mu a}p_\mu\otimes \hat t_a~.
\end{aligned}
\end{equation}
The homotopy Maurer-Cartan equations \eqref{eq:MCeqs} defining the various curvatures are then $-\CF=0$, where $\CF$ is the generalized 2-curvature \eqref{eq:generalized_2_curvature} as found before. Infinitesimal gauge transformations are parameterized by 
\begin{equation}
\lambda=\hat{A}^\alpha_r\otimes\hat\tau_\alpha -\hat{B}^a_{\mu r}\xi^\mu\otimes \hat t_a-\hat{B}^{\mu a}_{\phantom{\mu} r}\xi_\mu\otimes \hat t_a~,
\end{equation}
where $\hat{A}_r$ takes values in $W$ and $\hat{B}_{\mu r}$ and $\hat{B}^{\mu}_{\phantom{\mu} r}$ are $V[1]$-valued. Their general action \eqref{eq:MCgaugetrafos} amounts to \eqref{eq:infinit_gen_gauge_trafo} for a generalized 2-connection. Altogether, we recovered the gauge potential, the curvatures and the infinitesimal gauge transformations of generalized higher gauge theory.

\subsection{Global description}

Finite (small) gauge transformations can be obtained from the infinitesimal ones described above by using the integration method of \cite{Fiorenza:2010mh}, which follows an idea of \cite{Henriques:2006aa}. The local kinematical data can then be glued together on overlaps of patches of a cover by these finite gauge transformations. One disadvantage of this approach is the following. In certain simple cases as e.g.\ that of crossed modules of Lie algebras, there is a straightforward integration available, as e.g.\ that to a crossed module of Lie groups. The integration of \cite{Henriques:2006aa}, however, usually yields a different result, which is only categorically equivalent to that of the straightforward integration.

Here, we follow a slightly different route, starting from a description of the generalized principal 2-bundle without connection in terms of \v Cech cochains. Based on this description, the infinitesimal gauge transformations \eqref{eq:infinit_gen_gauge_trafo} are then readily integrated. We shall restrict ourselves to the case of strict Lie 2-algebras, which will simplify our discussion drastically.

First, recall that a crossed module of Lie groups $\sH\xrightarrow{~\dpar~}\sG$ is a pair of Lie groups $(\sH,\sG)$ together with a Lie group homomorphism $\dpar:\sH\rightarrow \sG$ as well as an action $\acton$ of $\sG$ on $\sH$ by automorphisms satisfying 
\begin{equation}
 \dpar(g\acton h)\ =\ g\dpar(h)g^{-1}\eand \dpar(h_1)\acton h_2\ =\ h_1h_2h_1^{-1}
\end{equation}
for all $g\in \sG$ and $h,h_1,h_2\in \sH$. The first equation is simply equivariance of $\dpar$, while the second relation is known as the Peiffer identity. Moreover, such a crossed module of Lie groups is categorically equivalent to a strict Lie 2-group\footnote{i.e.\ a monoidal category internal to the category of smooth manifolds in which the product is associative and unital and in which objects and morphisms are (strictly) invertible}  as follows, cf.\ \cite{Baez:0307200}. The underlying category is given by the groupoid $\sG\ltimes \sH\rightrightarrows \sG$ with structure maps
\begin{subequations}\label{eq:2-group-maps}
\begin{equation}
 \sfs(g,h)\coloneqq g~,~~~\sft(g,h)\coloneqq \dpar(h)g~,~~~\id(g)\coloneqq (g,\unit_\sH)~,~~~(\dpar(h)g,h')\circ (g,h)=(g,h'h)~.
\end{equation}
The monoidal product is given by 
\begin{equation}
 g\otimes g'\coloneqq gg'\eand (g,h)\otimes (g',h')\coloneqq (gg',h(g\acton h'))~,
\end{equation}
\end{subequations}
where $g,g'\in \sG$ and $h,h'\in \sH$.

Using the Courant algebroid $T^*[2]T[1]M$ in the description of gauge theory can be regarded as replacing a manifold $M$ by the categorified manifold or {\em 2-space} $T^*M\rightrightarrows M$. Recall that a {\em 2-space} $\CX$ is a category internal to $\CatMan$ and therefore consists of a manifold $\CX_0$ of objects, a manifold $\CX_1$ of morphisms as well as smooth maps $\sfs,\sft:\CX_1\rightrightarrows \CX_0$ and $\id:\CX_0\rightarrow \CX_1$ as well as a composition map $\circ:\CX_1\times_{\CX_0}\CX_1$ such that the usual axioms for the structure map in a category are satisfied, cf.\ \cite{Baez:2004in}. In the case of $T^*M\rightrightarrows M$, we have the structure maps\footnote{Such a category in which source and target maps agree, is called {\em skeletal}. This property will simplify some aspects of the subsequent discussion.}
\begin{equation}
 \sfs(x,p)\coloneqq x~,~~~\sft(x,p)\coloneqq x~,~~~\id(x)\coloneqq (x,0)~,~~~(x,p)\circ (x,q)=(x,p+q)~,
\end{equation}
where $x\in M$, $p,q\in T^*_xM$. Given a cover $U=\sqcup_a U_a$ of $M$, we have an obvious induced 2-cover by the 2-space $T^* U\rightrightarrows U$. This 2-cover gives rise to a \v Cech double groupoid
\begin{equation}
 \check{\CC}\coloneqq (T^*U^{[2]}\rightrightarrows U^{[2]})~\rightrightarrows~(T^*U \rightrightarrows U)~,
\end{equation}
where $U^{[2]}\coloneqq \sqcup_{a,b} U_a\cap U_b$. 

Given a crossed module of Lie groups, $\sH\xrightarrow{~\dpar~}\sG$, there is also a natural double groupoid corresponding to the delooping of the strict Lie 2-group $\CG=(\sG\ltimes \sH\rightrightarrows \sG)$ as a double groupoid
\begin{equation}
 \CB\CG\coloneqq (\sG\ltimes \sH\rightrightarrows \sG)~\rightrightarrows~(*\rightrightarrows *)~,
\end{equation}
where $*$ denotes the one-element or singleton set.

A ``generalized'' principal 2-bundle on the 2-space $T^*M\rightrightarrows M$ is naturally defined as a lax (double) functor from $\check \CC$ to $\CB\CG$, see e.g.\ \cite{Jurco:2014mva} for a very detailed related discussion. Explicitly, such a lax double functor consists of an ordinary functor
\begin{equation}
 (\Phi_{ab}):(T^*U^{[2]}\rightrightarrows U^{[2]})\rightarrow (\sG\ltimes \sH\rightrightarrows \sG)
\end{equation}
together with a double natural isomorphism
\begin{equation}
 (\Xi_{abc}):(\Phi_{ab}\otimes \Phi_{bc})\Rightarrow (\Phi_{ac})~,
\end{equation}
such that
\begin{equation}
 \Xi_{acd}\circ (\Xi_{abc}\otimes \id_{\Phi_{cd}})=\Xi_{abd}\circ (\id_{\Phi_{ab}}\otimes \Xi_{bcd})~.
\end{equation}
The ordinary functor $(\Phi_{ab})$ is encoded in maps $(g_{ab},h_{ab}):T^*U^{[2]}\rightarrow \sG\ltimes \sH$, where necessarily $(g_{ab})$ is independent of the fibers in $T^*U^{[2]}$ and $\dpar(h_{ab})=\unit_\sG$. The double natural isomorphism $(\Xi_{abc})$ gives rise to maps $(g_{abc},h_{abc}):U^{[3]}\rightarrow \sG\ltimes \sH$ with $U^{[3]}\coloneqq \sqcup_{a,b,c} U_a\cap U_b\cap U_c$, where the map $(g_{abc})$ is fully fixed by the source of the double natural isomorphism. The fact that $(\Phi_{ab})$ and $(\Xi_{abc})$ form a lax double functor yields the equations
\begin{equation}
\begin{aligned}
 &(g_{ac}(x),h_{ac}(x,p_1+p_2))\circ (g_{abc}(x),h_{abc}(x))=\\
 &\hspace{1cm}(g_{abc}(x),h_{abc}(x))\circ \Big((g_{ab}(x),h_{ab}(x,p_1))\otimes (g_{bc}(x),h_{bc}(x,p_2))\Big)~,
\end{aligned}
\end{equation}
where $x\in U^{[2]}$ and $p_{1,2}\in T^*U^{[2]}$. Using relations \eqref{eq:2-group-maps}, this is readily translated into
\begin{equation}\label{eq:cocycle_relations}
 \begin{aligned}
  g_{abc}(x)&~=~\dpar(h^{-1}_{abc}(x))~,\quad g_{ac}(x)=g_{ab}(x)g_{bc}(x)~,\\
  h_{ac}(x,p_1+p_2)~h_{abc}(x)&~=~h_{abc}(x)~h_{ab}(x,p_1)~(g_{ab}(x)\acton h_{bc}(x,p_2))~,\\
  h_{acd}(x)~h_{abc}(x)&~=~h_{abd}(x)~(g_{ab}(x)\acton h_{bcd}(x))~.
 \end{aligned}
\end{equation}
We will refer to a set of maps $(g_{ab},h_{ab},h_{abc})$ satisfying the above equations as generalized 1-cocycles.

As a consistency check of our derivation, we can imagine replacing the 2-space $T^*M\rightrightarrows M$ by a discrete 2-space $M\rightrightarrows M$. In this case, the cocycle relations \eqref{eq:cocycle_relations} restrict to the usual ones of a principal 2-bundle with strict structure 2-group as found e.g.\ in \cite{Baez:2004in}. Also, note that the cocycle relations \eqref{eq:cocycle_relations} were also derived in \cite{Baez:2004in} using a different approach.

Analogously, we can now derive coboundary relations as double natural isomorphisms between the lax double functors. These give rise to isomorphism classes of generalized principal 2-bundles. Such a cocycle consists of maps $(\gamma_a,\chi_a):T^*U\rightarrow \sG\ltimes \sH$, where $(\gamma_a)$ is independent of the fibers in $T^*U$ and $\dpar(\chi_{a})=\unit_\sG$ as well as maps $(\gamma_{ab},\chi_{ab}):U^{[2]}\rightarrow \sG\ltimes \sH$, where $(\gamma_{ab})$ is again fully fixed. Instead of listing the general coboundary relations, let us just state that a trivial generalized principal 2-bundle has generalized 1-cocycles
\begin{equation}\label{eq:trivial_coboundaries}
 \begin{aligned}
  g_{ab}(x)~&=~\dpar(h^{-1}_{abc}(x))\gamma_a \gamma_b^{-1}~,\\
  h_{ab}(x,p_1)~&=~\chi_{ab}(x)^{-1}~\chi_{a}(x,p_1)~\chi_{ab}(x)~(\gamma_{b}(x)\acton \chi^{-1}_{b}(x,0))~,\\
  h_{abc}(x)~&~=~\chi^{-1}_{ac}(x)~\chi_{ab}(x)~(g_{ab}(x)\acton \chi_{bc}(x))~.
 \end{aligned}
\end{equation}

To endow the principal 2-bundle with connection, we now put local kinematic data of generalized higher gauge theory as discussed above on each patch. On overlaps of patches $U_a\cap U_b$, the components are then glued together via gauge transformations. The latter arise from integrating the infinitesimal gauge transformations \eqref{eq:infinit_gen_gauge_trafo} as done in \cite{Demessie:2014ewa} and lead to
\begin{subequations}\label{eq:glue_together}
\begin{equation}
 \begin{aligned}
  A_{b,\mu}\xi^\mu&=\left(g_{ab}^{-1} A_{a,\mu}g_{ab}+g_{ab}^{-1}\partial_\mu g_{ab}+\mu_1(\Lambda_{ab,\mu})\right)\xi^\mu~,\\
  A^\mu_b\xi_\mu&=\left(g_{ab}^{-1} \tilde A^\mu_a g_{ab}+\mu_1(\Lambda^\mu_{ab})\right)\xi_\mu~,\\
 \end{aligned}
\end{equation}
and
\begin{equation}
 \begin{aligned}
  B_{b,\mu\nu}\xi^\mu\xi^\nu&=\left(-g_{ab}^{-1}\acton B_{a,\mu\nu}-2A_{b,\mu}\acton \Lambda_{ab,\nu}-2\der{x^\mu} \Lambda_{ab,\nu}-\mu_2(\Lambda_{ab,\mu},\Lambda_{ab,\nu})\right)\xi^\mu\xi^\nu~,\\
  B^{\mu\nu}_b\xi_\mu\xi_\nu&=\left(-g_{ab}^{-1}\acton \tilde B^{\mu\nu}_a-A^\mu_b\acton \Lambda^\nu_{ab}- \mu_2( \Lambda^\mu_{ab},\Lambda^\nu_{ab})\right)\xi_\mu\xi_\nu~,\\
  B_{b,\mu}{}^\nu\xi^\mu\xi_\nu&=\left(-g_{ab}^{-1}\acton B_{a,\mu}{}^\nu-A_{b,\mu}\acton \Lambda^\nu_{ab}-A^\nu_b\acton \Lambda_{ab,\mu}+\phantom{\der{x^\mu}}\right.\\ 
  &\hspace{6cm}\left.-\der{x^\mu}\Lambda^\nu_{ab}-\mu_2(\Lambda_{ab,\mu},\Lambda^\nu_{ab})\right)\xi^\mu\xi_\nu~,\\
  B^\mu_b p_\mu&=\left(g_{ab}^{-1}\acton B^\mu_a+\Lambda^\mu_{ab}\right)p_\mu~.
 \end{aligned}
\end{equation}
Here, the $g_{ab}$ are part of the generalized 1-cocycle and the $(\Lambda_{ab,\mu}\xi^\mu)$ and $(\Lambda^\mu_{ab}\xi_\mu)$ are additional $\sLie(\sH)$-valued 1-forms and vector fields on $U^{[2]}$ satisfying the cocycle condition
\begin{equation}
 \begin{aligned}
  \Lambda_{ac,\mu}\xi^\mu&=\left(\Lambda_{bc,\mu}+g^{-1}_{bc}\acton \Lambda_{ab,\mu}-g^{-1}_{ac}\acton(h_{abc}(\nabla_\mu\acton h_{abc}^{-1}))\right)\xi^\mu~,\\
  \Lambda^\mu_{ac}\xi_\mu&=\left(\Lambda^\mu_{bc}+g^{-1}_{bc}\acton\Lambda^\mu_{ab}-g^{-1}_{ac}\acton(h_{abc}(A^\mu\acton h_{abc}^{-1}))\right)\xi_\mu~,
 \end{aligned}
\end{equation}
\end{subequations}
where $\nabla_\mu\acton h\coloneqq\der{x^\mu}h+A_\mu\acton h$. Finite gauge transformations are also readily read off from \eqref{eq:glue_together}.

Recall that principal bundles with connection and their isomorphisms are captured by the first (non-abelian) Deligne cohomology class. Our formulas for generalized principal 2-bundles with 2-connections and their gauge transformations thus gives a complete description of the first generalized (non-abelian) Deligne cohomology class.

\section{Dynamics}

Having fixed the kinematical background, we are now in a position to consider dynamical principles. 

\subsection{Weil algebra and higher Chern-Simons theory}

A very natural action functional arising directly within the framework of N$Q$-manifolds is that of (higher) Chern-Simons theory, constructed via the AKSZ mechanism. The
methods from the original paper \cite{Alexandrov:1995kv} can be easily
generalized and applied to the higher, $L_\infty$-algebra, scenario, see e.g.\ \cite{Bojowald:0406445,Kotov:2007nr,Sati:0801.3480,Fiorenza:2011jr,Gruetzmann:2014ica}. In the
following we give a quick review of the necessary tools for ordinary higher gauge theory, referring to
the references for any further details.

Recall from section \ref{ssec:ordinary_gauge} that the map
$a:\,T[1]M\rightarrow \frg[1]$ is not a morphism of N$Q$-manifolds. It can,
however, be lifted to a map $f:\,T[1]M\rightarrow T[1]\frg[1]$, such that
\begin{equation}
  f^*(Z^A):=a^*(Z^A)~,\qquad
  f^*(\dd_\frg Z^A):=\left(Q_{T[1]M}\circ a^*-a^*\circ Q_\frg\right)(Z^A)~,
\end{equation}
where $(Z^A)$ are coordinates on $\frg[1]$ and $\dd_\frg$ is just the exterior differential on $\frg[1]$ of weight 1. Note that we have the homological vector field
\begin{equation}
 Q_{W}:=Q_{\frg}+\dd_\frg
\end{equation}
on $T[1]\frg[1]$. It acts on coordinates of $T[1]\frg[1]$ as follows:
\begin{equation}
 Q_W Z^A=Q_{\frg} Z^A+\dd_\frg Z^A\eand
 Q_W\dd_\frg Z^A=-\dd_\frg Q_{\frg}(Z^A)~.
\end{equation}
With respect to $Q_W$, the map $f$ is now indeed a morphism of N$Q$-manifolds. The algebra of functions $\CC^\infty(T[1]\frg[1])\cong \sSym(\frg[1]^*\oplus \frg[2]^*)$ together with $Q_W$ is known as the {\em Weil algebra} $W(\frg[1])$ of $\frg[1]$ and for $\frg$ an ordinary Lie algebra, this reproduces the
conventional definition of the Weil algebra. 

It is reasonable to assume that we are interested in objects that are invariant under
the action of the gauge $L_\infty$-algebra $\mathfrak{g}$: these are called
\textit{invariant polynomials}, inv$(\mathfrak{g}[1])$, and they are
described as follows. The Weil algebra fits into the sequence
\begin{equation}
 {\rm inv}(\frg[1]) \ \embd\ \Walg(\frg[1]) \ \xrightarrow{\pi_\Walg} \ \CEalg(\frg[1])~,
\end{equation}
where $\pi_\Walg$ is the obvious projection by pulling back along the embedding $\frg[1]\embd T[1]\frg[1]$ as zero sections of the vector bundle. The invariant polynomials inv$(\frg[1])$ are then elements in $W(\frg[1])$ that sit completely in
$\sSym(\frg^*[2])$ and are closed under $Q_\Walg$. In other words,
for $p\in$ inv$(\frg[1])$, we have that $\pi_\Walg(p)=0$ and, using $\dd_\frg^2=0$, $Q_Wp=Q_{\frg}
p=0$. It is clear that contraction with a generic
element $X\in\frg[1]$ vanishes, so that also $\mathcal{L}_X p=0$. It
is therefore these types of objects that are relevant for
constructing topological invariants or even physical models.

In the case of the AKSZ mechanism, we are interested in the invariant
polynomial corresponding to the symplectic structure on $\frg[1]$,
that is
$\omega=\tfrac12\dd_\frg Z^A \wedge\omega_{AB}\dd_\frg Z^B$. This symplectic structure has a local symplectic potential
\begin{equation}
 \alpha=\iota_\varepsilon\omega=\sum_A|Z^A|Z^A\iota_{\dpar_A}\omega~,\quad\dd_\frg\left(\tfrac{1}{n+1}\alpha\right)=\omega~,
\end{equation}
where $\varepsilon=\sum_A|Z^A|Z^A\iota_{\partial_A}$ is
the Euler vector field, $\dpar_A\coloneqq \der{Z^A}$ and
$|Z^A|$ indicates the degree of the coordinate
$Z^A$. We saw above that invariant polynomials on $T[1]\frg[1]$ have
to be of the form
$p=\sum_kp_{\alpha_1\cdots\alpha_k}\dd_\frg\xi^{\alpha_1}\wedge\cdots\wedge\dd_\frg\xi^{\alpha_k}$,
so that the invariant part of the lift of $\omega$ to the bundle is
$\hat\omega=\tfrac12\dd_\frg\xi^\alpha\wedge \hat\omega_{\alpha\beta}\dd_\frg\xi^\beta$. One
can therefore ask whether an object $\chi$ exists, such that
$Q_\Walg\chi=\hat\omega$, and the projection to the
Chevalley-Eilenberg cohomology on $\frg[1]$ gives a cocycle $\kappa$
on $\CEalg(\frg[1])$, i.e.\ $\chi|_\CEalg(\frg[1])=\kappa$ with
$Q_{\frg}\kappa=0$. Such an object $\chi$ is called a {\em transgression element}. It allows to map the cohomology of $T[1]\frg[1]$ onto that of $T[1]M$: indeed, since $Q_{T[1]M}\circ f^*=f^*\circ
Q_\Walg$, exact objects in $T[1]\frg[1]$ get pulled back to exact
objects in $T[1]M$, in particular $\dd f^*(\chi)=f^*(\hat\omega)$.
  
A particularly interesting cocycle is the Hamiltonian $\CS$ of $Q_{\frg}$, which satisfies 
\begin{equation}
  Q_{\frg}\psi=\{\CS,\psi\}
\end{equation}
for any $\psi\in\CC^\infty(\frg[1])$, where $\{-,-\}$ is the Poisson bracket induced by the symplectic structure $\omega$ and the relation $Q^2=0$ amounts to $\{\CS,\CS\}=0$. The transgression element for this cocycle will be called a {\em Chern-Simons element}, and can be found in the following way. Starting with the
lift of $\alpha$ to the tangent bundle,
$\hat\alpha=\sum_B|Z^B|Z^B\omega_{BC}Q_\Walg Z^C$,
we see by how much $Q_\Walg\hat\alpha$ fails to be in ${\rm inv}(\frg[1])$:
\begin{equation}
\begin{aligned}
Q_\Walg\hat\alpha=&Q_{\frg}\hat\alpha+\dd_\frg\hat\alpha=\tfrac{1}{2}\sum_B|Z^B|\omega_{BC}\left(Q_{\frg}^B+\dd_\frg Z^B\right)\left(Q_{\frg}^C+\dd_\frg Z^C\right)\\
=&(n+1)\hat\omega+\sum_B|Z^B|\omega_{BC}Q_{\frg}^B\dd_\frg Z^C~.
\end{aligned}
\end{equation}
Combining this with
\begin{equation}
 Q_\Walg
 \CS=\dd_\frg \CS =\dd_\frg\left(\tfrac{1}{n+2}\iota_{Q_{\frg}}\iota_\epsilon\omega\right)=\sum_B|Z^B|\omega_{BC}Q_{\frg}^B\dd_\frg Z^C~,
\end{equation}
where in the first equality we used that
$\{\CS,\CS\}=Q_{\frg}\CS=0$, we obtain the Chern-Simons element as
\begin{equation}
  \chi=\tfrac{1}{n+1}\left(\hat\alpha-\CS\right)~.
\end{equation}

The Lagrangian for {\em higher Chern-Simons theory}\footnote{A related route to higher Chern-Simons theory was followed in \cite{Soncini:2014ara}, see also \cite{Zucchini:2015ohw}.} is now simply the pullback of the Chern-Simons element of the gauge $L_\infty$-algebra along $f$:
\begin{equation}\label{eq:higher_chern_simons_action}
  S_{\text{CS}}=\int_M f^*(\chi)~,
\end{equation}
where we identified polynomials in the $Z^A$ of degree $n$ with $n$-forms. The field content consists of an $n$-connection encoded in the morphism of N$Q$-manifolds $a$ which was lifted to $f$. The equations of motion of \eqref{eq:higher_chern_simons_action} are simply the homotopy Maurer-Cartan equations \eqref{eq:MCeqs} yielding a flat higher connection. For details on such models see again \cite{Ritter:2015ymv} and references therein.

\subsection{Generalized higher Chern-Simons theory}
Let us now apply the AKSZ construction to obtain the Chern-Simons form of generalized higher gauge theory. As discussed in the previous
section, we will have to pull back the Chern-Simons element on a Lie
2-algebra $\frg=W\leftarrow V[1]$ to the N$Q$-manifold $T^*[2]T[1]\FR^4$ along
\begin{equation}
  f:\, T^*[2]T[1]\FR^4\longrightarrow T[1]\frg[1]~,
\end{equation}
which is a lift of the map $a:\,T^*[2]T[1]\FR^4\longrightarrow
\frg[1]$. Note that in this setup, the gauge connection has significantly more components than in ordinary higher gauge theory.
Recall from section \ref{subsec:localdblhg} that a generalized 2-connection is of the form
\begin{equation}
 \CA=A+B=A_\mu\xi^\mu+A^\mu \xi_\mu+\tfrac12 B_{\mu\nu}\xi^\mu\xi^\nu+B_{\mu}{}^\nu\xi^\mu\xi_\nu+\tfrac12 B^{\mu\nu}\xi_\mu\xi_\nu + B^{\mu}p_\mu~,
\end{equation}
where $A$ takes values in $W$, $B$ is $V[1]$-valued and we used again coordinates $(x^\mu,\xi^\mu,\xi_\mu,$ $p_\mu)$ of degree $(0, 1,1,2)$, respectively, on $\CV_2$. We will work with the general twisted homological vector field
\begin{equation}
 \tilde Q_{\CV_2}=\xi^\mu\der{x^\mu}+p_\mu\der{\xi_\mu}+\frac12 T_{\mu\nu\kappa}\xi^\mu\xi^\nu\der{\xi_\kappa}+\frac{1}{3!}\der{x^\mu}T_{\nu\kappa\lambda}\xi^\nu\xi^\kappa\xi^\lambda\der{p_\mu}~,
\end{equation}
where $T_{\mu\nu\kappa}$ are the components of a closed 3-form on $M$.

On the Lie 2-algebra $\frg[1]=W[1]\leftarrow V[2]$ in coordinates
$(w^\alpha,\,v^a)$ of degrees $(1, 2)$, respectively, the symplectic structure is given by
$\omega=\dd v^a \wedge\omega_{a\alpha}\dd w^\alpha$ and a symplectic potential for $\omega$ is
$\alpha=2v^a\omega_{a\alpha}\dd w^\alpha+w^\alpha\omega_{a\alpha}\dd v^a$. Since $\omega$ has to be non-degenerate, we necessarily have $\dim W=\dim V$. As usual, this symplectic structure induces an inner product $(-,-)$ on $\frg$ with $(\hat\tau_\alpha,\hat t_a)=(\hat t_a,\hat\tau_\alpha):=\omega_{\alpha a}$, where $(\hat \tau_\alpha)$ and $(\hat t_a)$ are bases for $W$ and $V[1]$.

The sign conventions for the $Q$-structure are chosen as follows:
\begin{equation}
  Q_{\mathfrak g}=\left(-\frac12 m^\alpha_{\beta\gamma}w^\beta w^\gamma - m^\alpha_a
    v^a\right)\der{w^\alpha}
    +\left(\frac{1}{3!}m_{\alpha\beta\gamma}^a w^\alpha w^\beta w^\gamma -m_{\alpha b}^a w^\alpha
      v^b\right)\der{v^a}~,
\end{equation}
so that we get the corresponding Hamiltonian
\begin{equation}
  \mathcal{S}=\omega_{\alpha a}\left(-\frac12 m^\alpha_{\beta\gamma}w^\beta w^\gamma v^a -\frac12 m^\alpha_b
    v^a v^b\right)
    +\omega_{a\alpha}\left(\frac{1}{4!}m_{\delta\beta\gamma}^aw^\delta w^\beta w^\gamma w^\alpha -m_{\delta b}^aw^\delta
      v^b w^\alpha\right)~.
\end{equation}

The Chern-Simons element on this N$Q$-manifold is now readily calculated to be
\begin{equation}
 \tfrac13(\alpha-\CS)=\tfrac13(\omega_{a\alpha}(2v^a\dd w^\alpha+w^\alpha \dd v^a)-\mathcal{S})~.
\end{equation}
Its pullback along $f$ then yields the generalized higher Chern-Simons action:
\begin{align}\label{generalized higher CS action}
  S_{\rm ghCS}=&\int_{\FR^4} {\rm vol}~ f^*(\tfrac13(\alpha-\CS))\nonumber\\
=&\tfrac13\int_{\FR^4} {\rm vol}~ \Big(\big( 2B+A,
    \CF\big) -
    \tfrac12\big(\mu_2(A,A)
    +\mu_1(B),B\big)-\tfrac{1}{4!}\big(\mu_3(A,A,A),A\big)
   \Big)\nonumber\\
=&\int_{\FR^4} {\rm vol}~\left(\big(
   B,-\dd A+\tfrac12\mu_2(A,A)+\tfrac12\mu_1(B)\big)-\tfrac{1}{4!}\big(\mu_3(A,A,A),A\big)\right)
   + S_{\text{pT}}~,
\end{align}
where ${\rm vol}$ is the volume form on $\FR^4$ and $S_{\text{pT}}$ is a further contribution coming mostly from the twist term
$T$,
\begin{align}\label{SpH}
  S_{\text{pT}}=&\tfrac13\int_{\FR^4}{\rm vol}~\left( \big(
    A,\tfrac{1}{3!}\chi^\mu\partial_\mu T+(p_\nu+\tfrac12
    T_{\gamma\delta\nu}\xi^\gamma\xi^\delta)(B^{\nu r}\xi_r+B^\nu_{\phantom\nu
  \lambda}\xi^\lambda)\big)+    \right.\nonumber\\
&\hspace{4cm}\left.+2\big(B, (p_\nu+\tfrac12 T_{\gamma\delta\nu}\xi^\gamma\xi^\delta)A^\nu\big)\right)~.
\end{align}

Note that the action functional is a function on $T^*[2]T[1]\FR^4$ of degree 4. This is due to the following fact: In ordinary (higher) gauge theory where we use $T[1]M$, we can identify functions of degree 4 with the volume form on $\FR^4$. In generalized higher gauge theory, however, this identification is no longer possible, and one should integrate each component of the Lagrangian $f^*(\tfrac13(\alpha-\CS))$ separately.

The stationary points of the action functional $S_{\rm ghCS}$ are now given by totally flat generalized 2-connections. That is, the equations of motion simply read as $\CF=0$ with $\CF$ given in \eqref{eq:generalized_2_curvature}.

\subsection{3-Lie algebra based (2,0)-model}\label{ssec:2-0-model}

Another interesting application of generalized higher gauge theory is an interpretation of the effective M5-brane dynamics proposed in \cite{Lambert:2010wm}. In these equations, the field content consists of a six-dimensional (2,0)-supermultiplet and an additional vector field, both taking values in a 3-Lie algebra $\fra$, as well as a gauge potential taking values in the inner derivations of $\fra$.

To discuss the model, let us focus on the 3-Lie algebra $A_4$, which is readily interpreted as the semistrict Lie 2-algebra based on the N$Q$-manifold
\begin{equation}
 A_4~=~(~*\leftarrow \aso(4)[1]\leftarrow \FR^4[2]~)~.
\end{equation}
Introducing local coordinates $\chi^a$, $a=1,\ldots,4$ on $\FR^4$ and $\gamma^{ab}=-\gamma^{ba}$ on $\aso(4)$ with corresponding basis $\tau_{ab}=-\tau_{ba}$ and $t_a$, the homological vector field reads as
\begin{equation}
 Q_{A_4}=-\frac12(\eps_{abce}\delta_{df}+\eps_{abdf}\delta_{ce})\gamma^{ab}\gamma^{cd}\der{\gamma^{ef}}-\eps_{abcd}\gamma^{ab}\chi^c\der{\chi^d}~.
\end{equation}
Here, the first summand in $Q_{A_4}$ encodes the Lie algebra of $\aso(4)$ in a convenient basis, while the second one describes the fundamental action of $\aso(4)$ on $\FR^4$. Note that the corresponding 2-term $L_\infty$-algebra with graded basis $(\hat \tau_{ab})$ and $(\hat{t}_a)$ has $\mu_1(\hat t_a)=0$. We can naturally endow this $L_\infty$-algebra with the Euclidean inner product $(\hat t_a,\hat t_b)\coloneqq \delta_{ab}$ on $\FR^4[1]$ as well as the split inner product $\lbr \hat \tau_{ab},\hat \tau_{bc}\rbr=\eps_{abcd}$ on $\aso(4)\cong \aso(3)\oplus \aso(3)$. Note that these inner products do {\em not} originate from a natural one via a symplectic structure on the N$Q$-manifold $A_4$. However, both induce a map $D:\FR^4\wedge\FR^4\rightarrow \aso(4)$ via
\begin{equation}
 \lbr y,D(\chi_1,\chi_2)\rbr\coloneqq (y\chi_1,\chi_2)=-(y\chi_2,\chi_1)
\end{equation}
for all $y\in \aso(4)$ and $\chi_{1,2}\in \FR^4$. Explicitly, we have $D(\hat t_a,\hat t_b)=\hat \tau_{ab}$. The totally antisymmetric, ternary bracket of the 3-Lie algebra here corresponds to
\begin{equation}
 [\chi_1,\chi_2,\chi_3]\coloneqq D(\chi_1,\chi_2)\chi_3~,
\end{equation}
implying $[\hat t_a,\hat t_b,\hat t_c]=\eps_{abcd}\hat t_d$. For more details, see \cite{Palmer:2012ya} and references therein.

Let us denote the bosonic and fermionic matter fields in the (2,0)-supermultiplet on $\FR^{1,5}$ by $X^I$ and $\Psi$, $I=1,\ldots,5$, respectively. In addition, we consider a 3-form $h=\tfrac{1}{3!}h_{\mu\nu\kappa}\dd x^\mu\wedge \dd x^\nu\wedge \dd x^\kappa$, $\mu,\nu,\kappa=0,\ldots,5$, as well as a vector field $C=C^\mu\der{x^\mu}$. All of these fields take values in $\FR^4$. We complement these fields by a gauge potential $A=A_\mu\dd x^\mu$ taking values in $\aso(4)$.  The full equations of motion proposed in \cite{Lambert:2010wm} then read as 
\begin{subequations}\label{eq:eomTensor}
\begin{eqnarray}
 \nabla^2 X^I-\tfrac{\di}{2}D(C^\nu,\bar{\Psi})\Gamma_\nu\Gamma^I\Psi+D(C^\nu,X^J)(D(C_\nu,X^J)X^I)&=&0~,\\
 \Gamma^\mu\nabla_\mu\Psi+D(C_\nu,X^I)\Gamma_\nu\Gamma^I\Psi&=&0~,\\
 \nabla_{[\mu}h_{\nu\kappa\lambda]}+\tfrac{1}{4}\eps_{\mu\nu\kappa\lambda\sigma\tau}D(C^\sigma,X^I)\nabla^\tau X^I+\tfrac{\di}{8}\eps_{\mu\nu\kappa\lambda\sigma\tau}D(C^\sigma,\bar{\Psi})\Gamma^\tau\Psi&=&0~,\\
h_{\mu\nu\kappa}-\tfrac{1}{3!}\eps_{\mu\nu\kappa\rho\sigma\tau}h^{\rho\sigma\tau}&=&0~,\\
F_{\mu\nu}-D(C^\lambda,h_{\mu\nu\lambda})&=&0~,\label{eq:eomTensor_e}\\
\nabla_\mu C^\nu=D(C^\mu,C^\nu)&=&0~\label{eq:eomTensor_f},\\
D(C^\rho,\nabla_r X^I)=D(C^\rho,\nabla_r\Psi)=D(C^\rho,\nabla_r h_{\mu\nu\lambda})&=&0~,
\end{eqnarray}
\end{subequations}
where $(\Gamma_\mu,\Gamma_I)$ are the generators of the Clifford algebra of $\FR^{1,10}$. We shall be interested only in the gauge part of these equations, captured by the field $h_{\mu\nu\kappa}$ and the gauge potential $A_\mu$.

Let us briefly summarize the analysis of equations \eqref{eq:eomTensor} as given in \cite{Palmer:2012ya}. The equation $D(C^\mu,C^\nu)=0$ implies that $C^\mu$ factorizes into the components $c^\mu$ of a vector field on $\FR^{1,5}$ and a constant element $v$ of $\FR^4[1]$: $C^\mu=c^\mu v$. Plugging this back into equation \eqref{eq:eomTensor_e}, we see that the curvature takes values in a subalgebra $\aso(3)$ of $\aso(4)$ generated by $D(v,-)$. The flatness in the other direction together with the non-abelian Poincar\'e lemma implies that we can choose a gauge such that $A$ also takes values in this subalgebra. As shown in appendix \ref{app:A}, the smaller N-manifold $*\leftarrow \aso(3)[1]\leftarrow \FR^4[2]$ underlies a strict Lie 2-algebra\footnote{which, however, is {\em not} a Lie 2-subalgebra of $A_4$} $A_4^v$ with $\mu_1(\hat t_a)\coloneqq D(v,\hat t_a)$.

One problem of equations \eqref{eq:eomTensor} is that the field $h_{\mu\nu\kappa}$ is {\em not} the curvature of a 2-form potential. Equation \eqref{eq:eomTensor_e} can either be regarded as being close to a transgression as done in \cite{Papageorgakis:2011xg}, or, as suggested in \cite{Palmer:2012ya} as the fake curvature equation in higher gauge theory. The latter suggests that we define a 2-form $B$ satisfying $c^\nu B_{\mu\nu}=0$ via
\begin{equation}\label{eq:Defh}
 h_{\mu\nu\kappa}=\frac{1}{|c|^2}\left(B_{[\mu\nu}c_{\kappa]}+\frac{1}{3!}\eps_{\mu\nu\kappa\lambda\rho\sigma}B^{[\lambda r}c^{\sigma]}\right)~.
\end{equation}
The resulting 3-form curvature is again self-dual:
\begin{equation}
 H=\dd B+\mu_2(A,B)=* H~.
\end{equation}
Replacing $h$ by $B$ as appropriate, we can now interpret the reformulated equations of motion within generalized higher gauge theory, using a generalized 2-connection $\CA$, cf.\ \eqref{eq:gen_2_connection}. The relevant Courant algebroid is $T^*[2]T[1]\FR^{1,5}$ and the Lie 2-algebra is $\aso(3)\leftarrow\FR^4[1]$. As dynamical principal, we will demand that most of the components of the generalized 2-curvature \eqref{eq:gen_2_curvature_components} vanish. Explicitly, we put the following components of $\CF$ to zero:
\begin{equation}
F^\mu p_\mu~,~~~H^{\mu\nu}\xi_\mu p_\nu~,~~~\tfrac12 H_{\mu\nu}{}^\kappa\xi^\mu\xi^\nu\xi_\kappa~,~~~H_\mu{}^\nu\xi^\mu p_\nu\eand\tfrac12 F_{\mu\nu}\xi^\mu\xi^\nu~. 
\end{equation}
Furthermore, the component $\tfrac{1}{3!}H_{\mu\nu\kappa}\xi^\mu\xi^\nu\xi^\kappa$ encodes a self-dual 3-form on $\FR^{1,5}$. Identifying the vector field $C^\mu$ with the component $B^\mu$ of the generalized 2-connection, we also impose the constraint $D(B^\mu,B^\nu)=0$. This implies the above described factorization $B^\mu=b^\mu v$ for some vector field $b^\mu$ on $\FR^{1,5}$ and $v$ the constant element in $\FR^4[1]$ defining $\mu_1$ in the Lie 2-algebra $A_4^v$ discussed in appendix \ref{app:A}. Since the component $F^\mu p_\mu$ of $\CF$ vanishes and $\mu_1(B^\mu)=0$, we conclude that the component $A^\mu\xi_\mu$ of $\CA$ vanishes. Considering the component $H^{\mu\nu}$, we learn that also the component $\tfrac12 B^{\mu\nu}\xi_\mu\xi_\nu$ of $\CA$ vanishes. Vanishing of $H_{\mu\nu}{}^\kappa$ implies that $B_\mu{}^\nu$ is covariantly constant, which implies that it can be gauged away. Finally, $H_\mu{}^\nu=0$ and $F_{\mu\nu}=0$ lead to equations \eqref{eq:eomTensor_e} and \eqref{eq:eomTensor_f}, respectively. The only remaining non-trivial component of $\CF$ is then indeed $H_{\mu\nu\kappa}$, which describes the self-dual 3-form. Altogether, we thus recover the gauge part of equations \eqref{eq:eomTensor}.

\section*{Acknowledgements}
This work was initiated during the workshop ``Higher Structures in String Theory and Quantum Field Theory'' at the Erwin Schr\"odinger International Institute for Mathematical Physics and we would like to thank the organizers and the institute for hospitality. The work of CS and LS was partially supported by the Consolidated Grant ST/L000334/1 from the UK Science and Technology Facilities Council.

\appendices

\subsection{The strict Lie 2-algebra of \texorpdfstring{$A^v_4$}{Av4}}\label{app:A}

We describe the 3-Lie algebra $A_4$ again as in section \ref{ssec:2-0-model}, using basis vectors $(\hat t_a)$ and $(\hat \tau_{ab})$ on $\FR^4[1]$ and $\aso(4)$, respectively. Without restriction of generality, we assume that the given vector $v\in A_4$ is aligned in a particular direction, $v=|v|\hat t_4$. Note that the subspace spanned by the $D(v,\hat t_a)$ forms the Lie subalgebra $\aso(3)\subset \aso(4)$, since
\begin{equation}
 [\hat \tau_{4i},\hat \tau_{4j}]=\eps_{4ijk}\hat \tau_{4k}~.
\end{equation}
We therefore have a 2-term sequence of vector spaces $\aso(3)\stackrel{\mu_1}{\longleftarrow}\FR^4[1]$ with 
\begin{equation}
 \mu_1(\hat t_a)=D(v,\hat t_a)=|v|\hat \tau_{4a}~.
\end{equation}
Together with the Lie bracket on $\aso(3)$,
\begin{equation}
 \mu_2(\hat \tau_{4i},\hat \tau_{4j})=\eps_{4ijk}\hat \tau_{4k}~,
\end{equation}
and the action of this subalgebra on $\FR^4[1]$ as induced from $A_4$,
\begin{equation}
 \mu_2(\hat \tau_{4i},\hat t_a)=\eps_{4iaj}\hat t_j~.
\end{equation}
form a strict Lie 2-algebra $A_4^v$, because
\begin{equation}
 \begin{aligned}
    \mu_1(\mu_2(\hat \tau_{4i},\hat t_a))&=|v|\eps_{4iaj}\hat \tau_{4j}=\mu_2(\hat \tau_{4i},\mu_1(\hat t_a))~,\\
    \mu_2(\mu_1(\hat t_a),\hat t_b)&=|v|\eps_{4abj}\hat t_j=\mu_2(\hat t_a,\mu_1(\hat t_b))~.
 \end{aligned}
\end{equation}
Altogether, the strict Lie 2-algebra $A_4$ with trivial $\mu_1$ is turned into the strict Lie 2-algebra $A^v_4$ with non-trivial $\mu_1$.

\bibliography{bigone}

\bibliographystyle{latexeu}

\end{document}